\documentclass[prd,showpacs,preprintnumbers,twocolumn,amsmath,nofootinbib,amssymb,floatfix]{revtex4}
\usepackage{graphicx,color,dcolumn,booktabs,bm}
\usepackage{epstopdf}
\usepackage{longtable,lscape}
\usepackage{txfonts}
\usepackage{overpic}
\usepackage{float}
\usepackage{amssymb}
\usepackage{amsmath}
\usepackage{epstopdf}
\usepackage{indentfirst}
\usepackage{feynmf}
\usepackage{slashed}
\usepackage{cases}
\usepackage{ulem}
\usepackage{color}
\usepackage{float}
\usepackage{multirow}
\usepackage{tikz}
\usepackage{epstopdf}
\usepackage{epsfig,dsfont,amssymb,amsmath,amsfonts,amsbsy,mathrsfs}
\usepackage{graphicx}
\usepackage{float}
\usepackage{subfigure}
\usepackage{array}
\usepackage{microtype}

\usepackage{float}
\usepackage{placeins}

\newcommand\orcid[1]{\href{https://orcid.org/#1}{\mbox{\scalerel*{\begin{tikzpicture}[yscale=-1,transform shape]\pic{orcidlogo};\end{tikzpicture}}{|}}}}

\begin{document}

\title{ Heavy-flavored meson spectrum under the action of external magnetic field}
\author{Bing-Rui Ma}
\affiliation{College of Physics and Electronic Information Engineering, Qinghai Normal University, Xining 810000, China}

\author{Hao Chen}
\email{chenhao_qhnu@outlook.com}
\affiliation{College of Physics and Electronic Information Engineering, Qinghai Normal University, Xining 810000, China}

\author{Cheng-Qun Pang}
\email{pcq@qhnu.edu.cn}
\affiliation{College of Physics and Electronic Information Engineering, Qinghai Normal University, Xining 810000, China}

\begin{abstract}
In this paper, the Schr{\"o}dinger equation in a magnetic field is utilized to study the effect of the magnetic field on $B$  mesons. The mass spetrum of $B$  mesons are numerically calculated for different magnetic field strengths by solving the Schr{\"o}dinger equation under the non-relativistic Cornell potential model, incorporating the Zeeman effect and the harmonic oscillator basis vector expansion method. The external magnetic field is assumed to be sufficiently strong in the calculations, so that the spin-orbit interaction energies can be omitted in comparison with the energies induced by the field. The Zeeman term in the Schr{\"o}dinger equation is taken to be  $ \frac{eB_s}{2m_ec}(m\pm1)$ where $m=0, \pm1$. The results demonstrate Zeeman splitting of energy levels in the external magnetic field.
\end{abstract}
\maketitle

\section{Introduction}
The beginning of the 20th century was an important transition from classical to modern physics. Over the past century, remarkable achievements have been made in understanding natural matter, both in the macroscopic and microscopic worlds\cite{Liao2012}. On the microscopic scale, experiments have evolved from atoms and nuclei to the level of quarks and leptons. Meanwhile, four fundamental interactions have been discovered: electromagnetic, weak, strong, and gravitational interactions. The study of particle physics has played an important role in the search for the nature of the natural world. In the study of particle physics, the study of heavy-flavor physics has played a very important role in testing and developing the Standard Model theory.

Nowadays, the bound states of heavy quarks with heavy quarks and heavy antiquarks have been extensively studied, and the non-relativistic potential model with second-order relativistic corrections has successfully explained the spectrum of heavy quarks and their decay properties. Physicists have discovered six flavors of quarks, which are $u$, $d$, $s$, $c$, $t$, and $b$, where $u$, $d$, and $s$ are light flavor quarks and the rest are heavy flavor quarks\cite{Guo2012}. The discovery of heavy quark symmetry and the formulation of the effective theory of heavy quarks have greatly improved the theoretical understanding of heavy quark systems. Based on the assumptions of the quark model, physicists consider mesons to be bound states consisting of quarks and antiquarks, and heavy-flavored mesons\cite{Liu2014}, i.e., hadronic states containing heavy-flavored quarks ($c$ quarks, $b$ quarks). Since the proposal of the component quark model, physicists have traditionally classified hadrons and regarded the basic structures of mesons and baryons as $q\overline{q}$ and $qqq$. In recent years, with the development of science and technology, many new heavy-flavored mesons have been discovered experimentally\cite{Guo2012}, such as $X(3872)$, $X(3940)$, $Y(4140)$, $Y(4260)$, and $Z^{*}(4430)$. These heavy-flavor mesons share the common feature of containing the structure of positive and negative quark pairs, but cannot be interpreted as conventional mesons or meson excited states . In order to study the structure and properties of these heavy-flavor mesons\cite{zhao2011, sun2012}, physicists have proposed a number of image-only quark models on the basis of quantum chromodynamics, including tetraquark states, molecular states, mixed states, and so on.

Currently, particle physicists have established the low mass spectra of the quark mesons $b\overline{b}$ and $c\overline{c}$, which provide very important experimental sites for the study of strong interactions. Similarly, the $B$ meson system is a large unexplored area\cite{LiZhen2015}. Since $B$ mesons usually decay into multiple particles through weak interactions, it is extremely difficult to explore $B$ mesons and their masses. Since the mid-1980s, the $B$ meson problem has been the most important area of research in particle physics, and its study requires new theoretical constructions. The $B$ meson is the bound state of a $b$ quark and a light antiquark, and the binding is provided by the strong interaction. In the last 50 years, with the rapid development of particle physics, the $B$ meson problem has become a popular area of high-energy physics as an important area of research to test the Standard Model and to find new physics\cite{Liu2017}.

\section{Mass spectrometry}
 The $B$ meson consists of a heavy quark and a light antiquark, and because of the large mass of the heavy quark, the velocity of the quarks in the $B$ meson system is relatively small. Therefore we can consider the $B$ meson system as a non-relativistic system, and the mass spectrum of the $B$ meson is obtained by solving the Schr {\"o}dinger equation, where the Hamiltonian is
\begin{equation}
\widetilde{H}=\sqrt{p+m_1^2}+\sqrt{p+m_2^2}+\widetilde{V}(p,r),
\end{equation}\\
where $m_1$ and $m_2$ are the masses of quarks and antiquarks, respectively, and in this paper the potential $\widetilde{V}$ consists a linear scalar potential in long-range region and a Coulomb-type vectorial potential via single-gluon exchange in short-range region.

In the non-relativistic limit, the $V(r)$ potential term is transformed into the familiar non-relativistic potential $V'(r)$, which is expressed as
\begin{equation}
V'(r)=H_{conf}+H_E,
\end{equation}
among them,
\begin{equation}
H_{conf}=\sigma r+c+(-\frac{4}{3}\frac{\alpha_s}{r})=V_S(r)+V_V(r),
\end{equation}
\begin{equation}
H_E=\frac{qB}{2\mu}(m\pm1).
\end{equation}

In Eq. $(3)$, $c$ is the tuning parameter as shown in TABLE \ref{tab_1} $1$, $\mu$ is the approximate mass, $\mu=\frac{m_1 m_2}{m_1+m_2}$ ,$\alpha_S$ is the coupling parameter, in this case it is the $H_{conf}$ term comprising the linearly bounded potential $V_V(r)=-\frac{4}{3}\frac{\alpha_{s}}{r}$ and the Coulomb potential $V_S(r)=\sigma r+c$; $H_E$ is the Zeeman term, and in this term the spin angular momentum is taken to be $(m + 1)$ when $S_z=\frac{1}{2}$ and $(m - 1)$ when $S_z=-\frac{1}{2}$.

\begin{table}[htbp]\small
     \caption{Parameters used to calculate the $B$ meson spectrum}
     \label{tab_1}
    \setlength{\abovecaptionskip}{0cm}
    \centering
    \vspace{-3mm}
    \begin{center}
    \setlength{\tabcolsep}{12mm}{
    \begin{tabular}{cc}
    \toprule[2pt]
    Parameters    &      Value\\
    \midrule[1pt]
   \small{ $m_u(m_d)$}    &    \small{0.35000 $GeV$}
   \vspace{1mm}\\
    \small{$m_b$}         &   \small{5.2468 $GeV$} \vspace{1mm}\\
    \small{$\alpha_s$}    &   \small{0.44379} \vspace{1mm}\\
    \small{$b$}           &      \small{0.17258} \vspace{1mm}\\
    \small{$\sigma$}      &      \small{1.9753}\vspace{1mm}\\
    \small{$c$}           &     \small{-0.83480}\vspace{1mm}\\
    \bottomrule[2pt]
    \end{tabular}}
    \end{center}
\end{table}

The mass spectrum and wave function of the $B$ meson can be obtained by solving the energy eigenvalues and eigenvectors of the Hamiltonian quantities in Eq. $(1)$, which can be expanded using the simple harmonic oscillator (SHO) basis. The SHO wave function\cite{Liu2011} has an explicit form in coordinate and momentum space, respectively:
\begin{equation}
    \Psi_{nLM_L}(r)=R_{nL}(r,\beta)Y_{LM_L}(\Omega_r),
\end{equation}
\begin{equation}
    \Psi_{nLM_L}(p)=R_{nL}(p,\beta)Y_{LM_L}(\Omega_p),
\end{equation}
where,
\begin{equation}
\begin{aligned}
   R_{nL}(r,\beta)&=
   \beta^\frac{3}{2}\sqrt{\frac{2n!}{\Gamma(n+L+3/2)}}\\&(\beta r)^L e^\frac{-r^2\beta^2}{2}
    {L_{n}^{L+1/2}}\!(\beta^2 r^2),
    \end{aligned}
\end{equation}
\begin{equation}
\begin{aligned}
R_{nL}(p,\beta)&=
    \frac{(-1)^n(-i)^L}{\beta^\frac{3}{2}}e^{-\frac{p^2}{2\beta^2}}\\
    &\sqrt{\frac{2n!}{\Gamma(n+L+3/2)}}(\frac{p}{\beta})^L{L_{n}^{L+1/2}}(\frac{p^2}{\beta^2}),
    \end{aligned}
\end{equation}
where $Y_{LM_L}(\Omega_r)$ is the spherical harmonic function, $R_{nl}(n=0,1,2,3,...) $ is the radial wave function and ${L_{n}^{L+1/2}}(x)$ denotes the Laguerre polynomial.

In order to obtain the mass spectrum of the $B$ meson, the corresponding parameters involved in the adopted potential model need to be determined by fitting the experimental data, as shown in TABLE \ref{tab_1}.

In the calculation process, the non-relativistic potential model (Cornell potential)\cite{Tan2003} is added, and the method of harmonic oscillator basis-vector expansion is applied. It is important to determine the values of each parameter in the equations as well as to unify their units, which are determined to be $GeV$ in this paper, applying the natural system of units. The rest masses of $b$ quarks and $u$ quarks are added to the kinetic energy term $-\frac{p^2}{2m}$, and the value of $\beta$ is taken to be the universal value of the meson, which is 0.5 \cite{Gao2003}, and the magnitude of the magnetic field in the Zeeman term $H_E$ is taken to be zero in units of $GeV$ in the computation of the mass spectrum of a $B$ meson with no magnetic field as is shown in TABLE \ref{tab_2}, in which the units of the final value have been converted to $MeV$.

 \begin{table}[htbp]\small
     \caption{Mass spectrum of a $B$ meson without magnetic field (unit: $MeV$)}
     \label{tab_2}
     \vspace{-3mm}
     \begin{center}
    \setlength{\abovecaptionskip}{0cm}
    \centering
    \setlength{\tabcolsep}{9mm}{
    \begin{tabular}{ccc}
    \toprule[2pt]
    $n^{2s+1}L_J$ & Expe. &  This work\\
    \midrule[1pt]
    $1^3 S_1$    &  5324   &  5413\vspace{1mm}\\
    $2^3 S_1$    &  -      &  6050\vspace{1mm}\\
    $1^1 P_1$    &  5726   &  5860 \vspace{1mm}\\
    $2^1 P_1$    &  -      &  6353 \vspace{1mm}\\
    \bottomrule[2pt]
    \end{tabular}}
    \end{center}
 \end{table}

\section{Mass spectra in the presence of an external magnetic field}
In the case of adding a magnetic field, we set the magnetic field strength $B$ to be $1 \sim 5$ $GeV$, and the charge $q$ is taken to be $-1$. For the $B$ meson, we calculate the $1^{3}S_1$, $2^{3}S_1$, $1^{1}P_1$ ,$2^{1}P_{1}$ states. The following values are calculated by taking $\frac{ e B_s}{2m_e c}(m-1)$ according to the $S_z=-\frac{1}{2}$ Zeeman term, and the units of the final values are transformed to $MeV$ as shown in TABLE \ref{tab_3}.

\begin{table*}[htbp]\small
  \caption{Mass spectrum of B meson in external magnetic field at $S_z=-\frac{1}{2}$ (unit: $MeV$)}
  \label{tab_3}
  \vspace{1mm}
  \setlength{\abovecaptionskip}{0cm}
  \centering
  \setlength{\tabcolsep}{5mm}
  \resizebox{0.7\linewidth}{!}{
\begin{tabular}{ccccccccccccc}
    \hline
    \toprule[2pt]
    $m-1$ & $n^{2s+1}L_J$ &  $B=1$ & $B=2$ & $B=3$ & $B=4$ & $B=5$\\
    \midrule[1pt]
    \multirow{4}*{$m=-1$} &  $1^3 S_1$  &  5414.77 & 5416.54 & 5418.31 & 5420.08 & 5421.85\vspace{1mm}\\
   \multirow{4}*{}        &  $2^3 S_1$  &  6051.77 & 6053.54 & 6055.31 & 6057.08 & 6058.85\vspace{1mm}\\
   \multirow{4}*{}        &  $1^1 P_1$  &  5861.77 & 5863.54 & 5865.31 & 5867.08 & 5868.85\vspace{1mm}\\
    \multirow{4}*{}       &  $2^1 P_1$  &  6354.77 & 6356.54 & 6358.31 & 6360.08 & 6361.85\vspace{1mm}\\
     \midrule[1pt]
     \multirow{4}*{$m=0$} &  $1^3 S_1$  &  5413.88 & 5414.77 & 5415.65 & 5416.54 & 5417.42\vspace{1mm}\\
   \multirow{4}*{}        &  $2^3 S_1$  &  6050.88 & 6051.77 & 6052.65 & 6053.54 & 6054.42\vspace{1mm}\\
   \multirow{4}*{}        & $1^1 P_1$   &  5860.88 & 5861.77 & 5862.65 & 5863.54 & 5864.42\vspace{1mm}\\
    \multirow{4}*{}       & $2^1 P_1$   &  6353.88 & 6354.77 & 6355.65 & 6356.54 & 6357.42\vspace{1mm}\\
    \midrule[1pt]
     \multirow{4}*{$m=1$} &  $1^3 S_1$  &  5413 & 5413 & 5413 & 5413 & 5413\vspace{1mm}\\
   \multirow{4}*{}        &  $2^3 S_1$  &  6050 & 6050 & 6050 & 6050 & 6050\vspace{1mm}\\
   \multirow{4}*{}        &  $1^1 P_1$  &  5860 & 5860 & 5860 & 5860 & 5860\vspace{1mm}\\
    \multirow{4}*{}       &  $2^1 P_1$  &  6353 & 6353 & 6353 & 6353 & 6353\vspace{1mm}\\
    \bottomrule[2pt]
    \hline
 \end{tabular}}
 \end{table*}

\begin{table*}[htbp]\small
  \caption{Mass spectrum of B meson in external magnetic field at $S_z=\frac{1}{2}$ (unit: $MeV$) }
  \label{tab_4}
  \vspace{-3mm}
  \begin{center}
  \setlength{\abovecaptionskip}{0cm}
  \centering
  \setlength{\tabcolsep}{5mm}
  \resizebox{0.7\linewidth}{!}{
  \begin{tabular}{ccccccccccccc}
  \hline
    \toprule[2pt]
    $m-1$ & $n^{2s+1}L_J$ &  $B=1$ & $B=2$ &$B=3$ & $B=4$ & $B=5$\\
    \midrule[1pt]
    \multirow{4}*{$m=0$} &  $1^3 S_1$  &  5412.12 & 5411.23 & 5410.35 & 5409.46 & 5408.58\vspace{1mm}\\
   \multirow{4}*{}       &  $2^3 S_1$  &  6049.12 & 6048.23 & 6047.35 & 6046.46 & 6045.58\vspace{1mm}\\
   \multirow{4}*{}       &  $1^1 P_1$  &  5859.12 & 5858.23 & 5857.35 & 5856.46 & 5855.58\vspace{1mm}\\
    \multirow{4}*{}      &  $2^1 P_1$  &  6352.12 & 6351.23 & 6350.35 & 6349.46 & 6348.58\vspace{1mm}\\
     \midrule[1pt]
     \multirow{4}*{$m=1$} &  $1^3 S_1$  &  5411.23 & 5409.46 & 5407.69 & 5405.92 & 5404.15\vspace{1mm}\\
   \multirow{4}*{}        &  $2^3 S_1$  & 6048.23  & 6046.46 & 6044.69 & 6042.92 & 6041.15\vspace{1mm}\\
   \multirow{4}*{}        &  $1^1 P_1$  &  5858.23 & 5856.46 & 5854.69 & 5852.92 & 5851.15\vspace{1mm}\\
    \multirow{4}*{}       &  $2^1 P_1$  &  6351.23 & 6349.46 & 6347.69 & 6345.92 & 6344.15\vspace{1mm}\\
    \midrule[1pt]
     \multirow{4}*{$m=-1$} &  $1^3 S_1$  &  5413 & 5413 & 5413 & 5413 & 5413\vspace{1mm}\\
   \multirow{4}*{}         &  $2^3 S_1$  &  6050 & 6050 & 6050 & 6050 & 6050\vspace{1mm}\\
   \multirow{4}*{}         &  $1^1 P_1$  &  5860 & 5860 & 5860 & 5860 & 5860\vspace{1mm}\\
    \multirow{4}*{}        &  $2^1 P_1$  &  6353 & 6353 & 6353 & 6353 & 6353\vspace{1mm}\\
    \bottomrule[2pt]
    \hline
 \end{tabular}}
 \end{center}
\end{table*}

 It is shown in TABLE \ref{tab_4}  that the results of the calculation based on $S_z=\frac{1}{2}$ with the Zeeman term taken as $\frac{ e B_s}{2m_e c}(m+1)$. In the calculation, replace $(m - 1)$ with $(m+1)$ in the Zeeman term.

\begin{figure*}[htbp]
    \centering
    \includegraphics[width=0.7\textwidth]{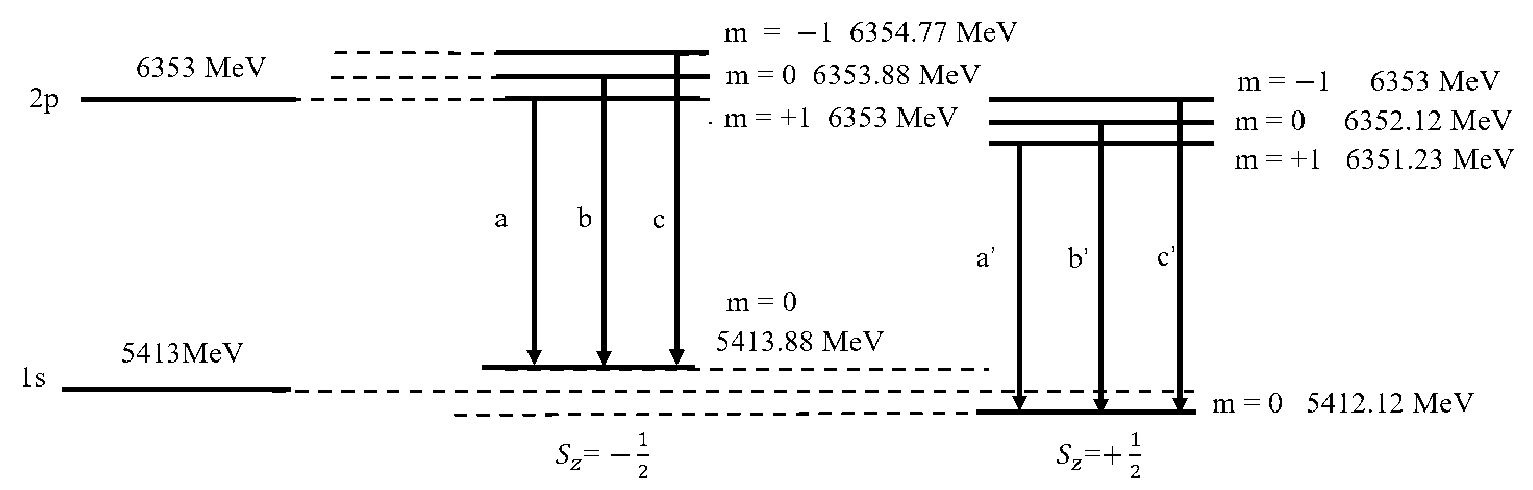}
    \caption{Splitting of the $B$ meson spectrum $s$ term and $p$ term in a magnetic field}
    \label{fig_1}
\end{figure*}

 The magnetic field strength $B=1 GeV$ is plotted through Tables \ref{tab_3} and \ref{tab_4}, and FIG. \ref{fig_1} represents the splitting of the $1s$ and $2p$ energy levels in the external magnetic field in units of $MeV$.

\begin{figure*}[htbp]
    \centering
  \includegraphics[width=0.7\textwidth]{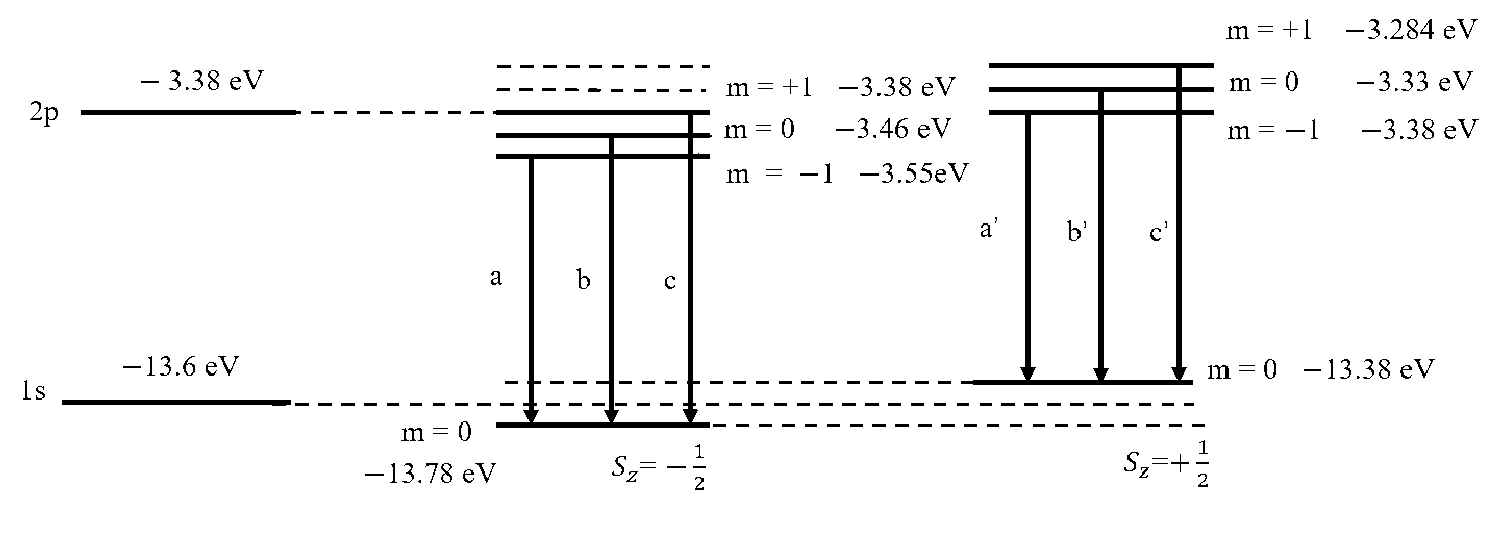}
 \caption{ Splitting of the $s$ and $p$ terms of the spectrum of a hydrogen atom in a magnetic field }
 \label{fig_2}
\end{figure*}

\section{Discussions}
In this paper, the $B$ meson spectrum is numerically calculated under the action of external magnetic field. In order to verify the accuracy of the program and the numerical value, the spectrum of hydrogen atom in the absence of a magnetic field and under the action of a magnetic field to verify the hydrogen atom spectrum has been calculated, the unit of each parameter in calculating the spectrum of hydrogen atom is $ keV$, and the unit of the final value is converted to $eV$. The hydrogen atom spectrum obtained by calculation is shown in TABLE \ref{tab_5}.

 \begin{table}[ht]\small
     \caption{Hydrogen atom spectrum (unit: $eV$)}
     \label{tab_5}
     \vspace{-3mm}
     \begin{center}
    \setlength{\abovecaptionskip}{0cm}
    \centering
    \setlength{\tabcolsep}{6mm}{
    \begin{tabular}{cccc}
    \toprule[2pt]
    $L$ & $\beta$ & Expe. & This work\\
    \midrule[1pt]
    $L=0$    &  $5.24$   &  $-13.6$ & $-13.580$\vspace{1mm}\\
    $L=1$    &  $2.53$   & $-3.39$  & $-3.388$\vspace{1mm}\\
    $L=2$    &  $1.05$   &  $-1.51$ &  $-1.511$ \vspace{1mm}\\
    $L=3$    &  $0.36$   & $-0.85$ &  $-0.850$ \vspace{1mm}\\
    $L=4$    &  $0.17$   & $-0.54$ & $-0.541$\vspace{1mm}\\
    \bottomrule[2pt]
    \end{tabular}}
    \end{center}
    \end{table}

From the calculated mass spectrum of the hydrogen atom, it can be seen that different states take different $\beta$ values, and the results calculated according to the corresponding $\beta$ values are similar to the experimental values in the quantum mechanics book\cite{RN10}. As the value of $\beta$ decreases, the mass of the hydrogen atom increases. In the following calculation of the hydrogen atom spectrum under the action of the magnetic field, the size of the magnetic field strength $B $ is set to $ 5 GeV $, and the corresponding $\beta $ value is kept unchanged. TABLE \ref{tab_6} shows the results calculated according to the $ S_z = -\frac{1}{2} $, the Zeeman term is taken as $ \frac{ e B_s}{2m_e c}(m-1)$.

    \begin{table}[ht]\small
     \caption{Hydrogen atom spectrum (unit: $eV$) under external magnetic field at $S_z=-\frac{1}{2}$.}
     \label{tab_6}
     \vspace{-3mm}
     \begin{center}
    \setlength{\abovecaptionskip}{0cm}
    \centering
    \setlength{\tabcolsep}{6mm}{
    \begin{tabular}{cccc}
    \toprule[2pt]
    $m+1$ & $L$ &  $\beta$ & This work\\
    \midrule[1pt]
    \multirow{2}*{$m=0$} & $L=0$ & 5.24  & $-13.786$ \vspace{1mm}\\~
   \multirow{2}*{} & $L=1$   &  2.53 & $-3.461$
\vspace{1mm}\\
    \midrule[1pt]
    \multirow{2}*{$m=-1$} &  $L=0$  &  5.24 & $-14.005$\vspace{1mm}\\
   \multirow{2}*{} & $L=1$   & 2.53  & $-3.555$
\vspace{1mm}\\
    \midrule[1pt]
     \multirow{2}*{$m=1$} &  $L=0$  &  5.24 & $-13.580$\vspace{1mm}\\
   \multirow{2}*{} & $L=1$   & 2.53 &  $-3.388$\vspace{1mm}\\
    \bottomrule[2pt]
    \end{tabular}}
    \end{center}
    \end{table}

TABLE \ref{tab_7} shows the results calculated based on $S_z=\frac{1}{2}$ with the Zeeman term taken as $\frac{ e B_s}{2m_e c}(m+1)$. In the calculation, replace $m- 1$ with $m + 1$ in the Zeeman term.

  \begin{table}[htbp]\small
     \caption{Hydrogen atom spectrum (unit: $eV$) under external magnetic field at $S_z=\frac{1}{2}$.}
     \label{tab_7}
      \vspace{-3mm}
     \begin{center}
    \setlength{\abovecaptionskip}{0cm}
    \centering
    \setlength{\tabcolsep}{6mm}{
    \begin{tabular}{cccc}
    \toprule[2pt]
    $m+1$ & $L$ &  $\beta$ & This work\\
    \midrule[1pt]
    \multirow{2}*{$m=0$} & $L=0$ & 5.24 & $-13.385$
\vspace{1mm}\\~
   \multirow{2}*{} & $L=1$   &  2.53 &  $-3.330$\vspace{1mm}\\
\midrule[1pt]
    \multirow{2}*{$m=-1$} &  $L=0$  &  5.24 & $-13.580$ \vspace{1mm}\\
   \multirow{2}*{} & $L=1$   & 2.53  & $-3.388$\vspace{1mm}\\
\midrule[1pt]
     \multirow{2}*{$m=1$} &  $L=0$  &  5.24 & $-13.201$\vspace{1mm}\\
   \multirow{2}*{} & $L=1$   & 2.53  & $-3.284$
\vspace{1mm}\\
    \bottomrule[2pt]
    \end{tabular}}
    \end{center}
\end{table}

According to the values in TABLE \ref{tab_5}, TABLE \ref{tab_6} and TABLE \ref{tab_7}, the splitting of the $1s$ and $2p$ energy levels in an external magnetic field in $eV$ is shown in FIG. \ref{fig_2}.

 The energy level splitting diagram drawn from the values of the spectrum of the hydrogen atom under the action of a magnetic field is the same as the diagram\cite{RN10} observed in the Stearns-Gerlach experiment, indicating that the calculated $B$ meson spectrum, as well as the $B$ meson spectrum under the action of a magnetic field are correct.

\section{summary}
The Schr{\"o}dinger equation is utilized to study the $B$ meson spectrum in a magnetic field. Under the non-relativistic Cornell potential model, the Zeeman effect and the harmonic oscillator basis expansion are applied to solve the Schr{\"o}dinger equation to obtain the mass spetrum of the $B$ meson in different magnetic fields, and the computational results show that when the Zeeman term is $\frac{ e B_s}{2m_e c}(m-1)$, and the magnetic quantum number $m$ is taken to be $0$ and $-1$, the mass of the $B$ meson increases with the increase of the magnetic field; when the magnetic quantum number $m$ is taken as $1$, the mass of the $B$ meson is the same as that in the absence of the magnetic field, regardless of the change of the magnetic field. When the Zeeman term is taken as $\frac{ e B_s}{2m_e c}(m+1)$ and the magnetic quantum number $m$ is taken as $0$ and $1$, the mass of the $B$ meson decreases with the increase of the magnetic field, and when the magnetic quantum number $m$ is taken as $-1$, the mass of the $B$ meson is the same as that in the absence of magnetic field, regardless of the change of the magnetic field.

In order to ensure the accuracy of the calculation process and the values, the same theory and method are used to calculate the hydrogen atom spectrum and the spectrum of hydrogen atom in the magnetic field. From the calculated hydrogen atom spectrum can be seen that different states take  different $\beta$ values, and the calculated results according to the corresponding $\beta$ value are similar to the experimental value. As the $\beta$ value decreases, the mass of the hydrogen atom increases. The energy level splitting diagrams drawn from the calculated spectrum of hydrogen atom in a magnetic field are the same as those observed in the experiment. This means that the calculated $B$ meson spectrum and the $B$ meson spectrum in the magnetic field are correct.

\bibliography{references}

\end{document}